# Solar Coronal Heating and Limb Effect

Yi-Jia Zheng

**National Astronomical Observatory, Chinese Academy of Sciences, Beijing, 100012 China**

**Abstarct**
According to the theory of the 'soft-photon process' proposed by Yijia Zheng in arxiv#1305.0427, the quiet solar coronal heating problem and the observed center-to-limb wavelength variations of the solar lines (limb effect) can be explained. In this paper the quantitative calculations for these two phenomena are presented.

## 1. Introduction

Solar corona is a hot outer layer of the sun where the temperature is a few million degrees, much higher than the effective temperature of the surface of the sun. The heating mechanism at work in the solar corona has puzzled researchers for more than six decades. Several heating mechanisms have been proposed. For a recent review see Klimchuk (2006).

Usually, except the scattering the solar corona is thought to be completely transparent for the non-scattered solar radiation. However, the solar corona is not completely transparent for the non-scattered solar radiation. As reported by Halm (1907) and subsequent observers, almost all absorption lines in the spectrum of the solar limb are shifted to the red as compared with their wavelengths at the center of the disk. This is the so-called solar limb effect. Zhao et al. (1998) investigated the formation of neutral magnesium emission lines in the solar photosphere in detail. They found that the observed profiles of selected lines can be reproduced well, but the center-to-limb variations of their center wavelength cannot be reproduced. These emission lines also show the solar limb effect. This means that the whole solar radiation were redshifted during the propagation in the solar corona. The redshift of the non-scattered solar radiation means a small part of their energy has losses during the propagation. The energy losses of the radiation are caused by soft photons released from the non-scattered radiation (soft-photon process proposed by Yijia Zheng in arxiv#1305.0427). The soft photons released are then absorbed by the solar corona and the corona is heated.

Withbroe and Noyes (1977) have estimated that an energy flux of $3\times10^5$ ergs s$^{-1}$ cm$^{-2}$ is required to maintain the quiet solar corona. The total solar radiation emitted at the solar surface is $6.27\times10^{10}$ ergs s$^{-1}$ cm$^{-2}$ (Allen 1973). Therefore, if the total solar radiation is redshifted by $5\times10^{-6}$ due to the energy losses, then the energy released by the non-scattered solar radiation due to the redshift is sufficient to heat the quiet solar corona.

Kierein and Sharp (1968) have shown that the variation in magnitude of the redshift from center to limb shows agreement with the variation in the number of electrons along the line of sight. They try to explain the redshift as the result of the Compton scattering interaction of photons and the electrons in the corona, but their explanation has encountered two difficulties. First, the redshift of the photon in the Compton scattering only occurs in the process in which the photons are scattered into another direction. If the photons propagate along the original direction, according to the Compton scattering theory, their frequency will not be changed. In the solar limb effect the observed photons have not been scattered, so the frequency of the photons should not be redshifted. Second, even the observed photons in the solar limb effect have been scattered by a small angle, the scattering cross section for a small angle Compton scattering is too small to explain the redshift integration effect. To show the solar limb effect they have to normalize the redshift at the disk center to 0.1 km s$^{-1}$. Therefore, the explanation of Kierein and Sharp is not acceptable.

Yijia Zheng (arxiv#1305.0427) proposed a new explanation on the origin of the redshift of the interaction between the non-scattered photons and electrons. The non-scattered photons should also interact with the plasma during the propagation and then emit the soft photons. Due to the emission of the soft photons, the non-scattered photons are redshifted. The observed redshift of the non-scattered photons is related to the number of electrons along the line of sight by

$$\ln(1+Z) = \frac{2e^2}{\pi.\eta}\sigma_c \int_0^R \frac{\ln \Lambda}{V_e} n_e(r)dr \qquad (1)$$

where $\sigma_c$ is the cross section of the Compton scattering, $e$ the charge of the electron, $\eta$ the Planck constant, $c$ the speed of light, $V_e$ the velocity of the electron, and $\Lambda$ a parameter determined by the plasma. Numerical calculation shows that in the solar corona the factor $\frac{2e^2}{\pi.\eta}\frac{\ln \Lambda}{V_e}$ is almost a constant. The variation of the factor in the solar atmosphere is discussed in section 3.

## 2. Model calculation

Referring to Figure 1, it can be seen that the path length of the observed light passing through the solar atmosphere varies depending on the position of emission. The path is least at the center and greatest at the solar limb. According to formula (1), the redshift observed is therefore greater at the limb than at the center. If the electron density distribution along the line of sight $n_e(L)$ in the solar corona is known, then using formula (1) the redshift caused by the solar corona can be calculated quantitatively. According to Allen (1973) the model of the quiet solar atmosphere can be tabulated as shown in Table 1.

Denoting the emitting source position on the solar disk by $\rho = \sin(\theta)$, then

$$L(n,\rho) = \sqrt{(R+h(n))^2 - (R\sin(\theta))^2} - R\cos(\theta) \qquad (2)$$

$$Z_n(\rho,n) \approx (n_e(n+1) + n_e(n))(L(n+1,\rho) - L(n,\rho))/2 \qquad (3)$$

$$Z(\rho) = \sum Z_n(\rho,n) \qquad (4)$$

The limb effect of redshift is
$$Z_{\lim b}(\rho) = Z(\rho) - Z(\rho = 0) \quad \dots\dots\dots\dots\dots\dots\dots\dots\dots\dots\dots\dots \quad (5)$$

Appenzeller and Schroter (1967) have measured the redshift of the continuum and several lines along polar and equatorial diameters of the solar disk. Fig. 2 shows their observed results for the core of $H_\beta$ and the results of the numerical calculation using formula (5). In the calculation the spectroheliogram height 1900 km for $H_\beta$ (Allen 1973) has been taken into account.

Table1. The smoothed electron density distribution $n_e(h)$ in the solar chromosphere and corona (Allen 1973)

| h (km) | Log $n_e$ (cm$^{-3}$) | Log n (cm$^{-3}$) | T |
|---|---|---|---|
| 0 | 11.96 | 16.13 | 4560 |
| 200 | 11.18 | 15.35 | 4180 |
| 500 | 10.88 | 14.08 | 5230 |
| 1000 | 10.87 | 12.25 | 6420 |
| 1500 | 10.54 | 11.17 | 8000 |
| 1900 | 10.49 | 10.82 | 11000 |
| 1990 | 10.10 | 10.40 | 28000 |
| 2000 | 9.81 | 10.11 | 100000 |
| 2010 | 9.47 | 9.77 | 190000 |
| 2100 | 9.02 | 9.32 | 470000 |
| 3500 | 8.65 | | |
| 7000 | 8.35 | | |
| 21000 | 8.18 | | |
| 42000 | 8.01 | | |
| 70000 | 7.89 | | |
| 140000 | 7.49 | | 1200000 |
| 280000 | 6.91 | | |
| 420000 | 6.48 | | |
| 560000 | 6.17 | | |
| 700000 | 5.91 | | 1800000 |

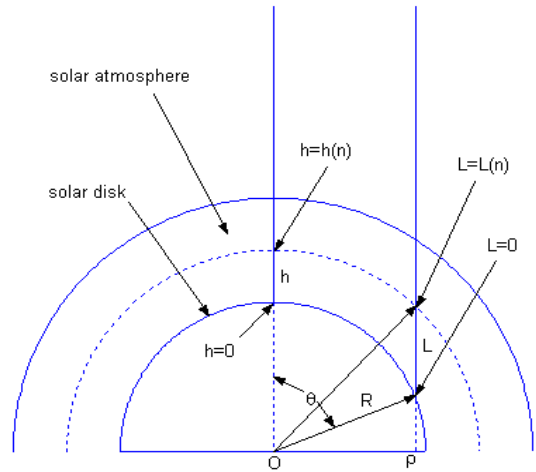

Fig 1. The line of sight in the solar atmosphere

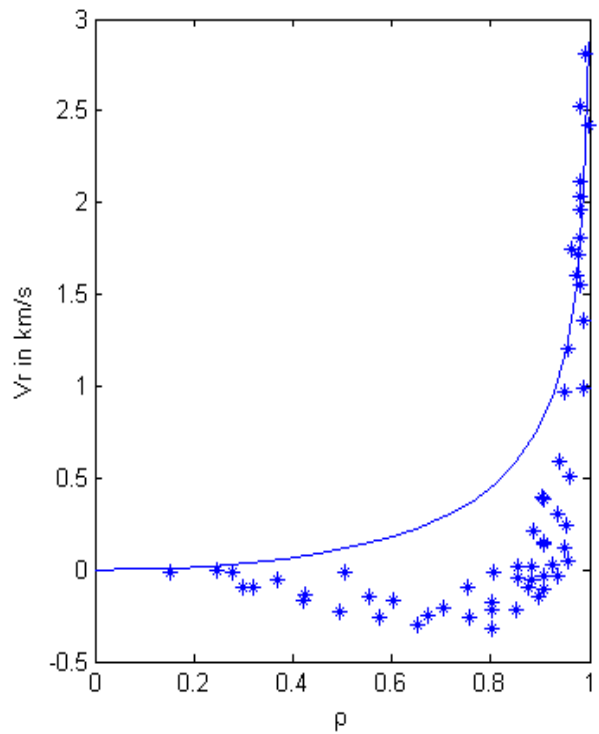

Fig2. The observed solar limb effect compare with the theory calculation. The observed data (asterisks) are taken from Appenzeller and Schroter 1967.

The energy released by the solar radiation due to the redshift can be calculated quantitatively now. The solar limb effect means that in the same layer of the solar atmosphere the solar radiation suffered from different redshifts in different directions. The energy released by the solar radiation in a layer can be calculated by

$$E_{released} = \int_0^{\pi/2} Z(n,\theta) \times F \times \cos(\theta) \times \sin(\theta) d\theta \int_0^{2\pi} d\phi \quad \ldots\ldots\ldots\ldots\ldots\ldots\ldots\ldots\ldots\ldots (6)$$

where $Z(n,\theta)$ can be calculated numerically using formula (3); $F$ is the mean radiation intensity of Sun's disk and equal to $2.0\times10^{10}$ erg cm$^{-2}$ s$^{-1}$ sr$^{-1}$ (Allen 1973).

Since the energy released by the solar radiation is taken off by the soft photons (Zheng, arxiv#1305.0427), the released energy is very easy to be absorbed by the plasma along the propagation path of the non-scattered photons, because the free-free linear absorption coefficient $\kappa_s$ is in proportion to $\frac{1}{v^3}$ (Allen 1973). The required input energy to compensate for the energy losses and the released energy caused by the redshift of the solar radiation for the solar corona and chromosphere layers are tabulated in table 2. The released energy caused by the redshift of the solar radiation for corona and chromosphere layers are calculated using formula (6). The required input energy is taken from Withbroe and Noyes (1977).

From Table 2 it can be seen that as a whole the energy released by solar radiation due to the redshift of the non-scattered solar radiation is sufficient to compensate for the energy losses. For the middle chromosphere the conductive energy input from other layers play an important role.

Table 2: The required input energy and the calculated released energy caused by the redshift of the solar radiation for corona and chromosphere layers. The required input energy is taken from Withbroe and Noyes (1977).

| Region name | Required input (erg s$^{-1}$ cm$^{-2}$) | Released energy due to redshift (erg s$^{-1}$ cm$^{-2}$) |
|---|---|---|
| Low chromosphere | $2\times10^6$ | $3.0\times10^6$ |
| Middle chromosphere | $2\times10^6$ | $7.2\times10^5$ |
| Upper chromosphere | $3\times10^5$ | $1.6\times10^6$ |
| Corona | $3\times10^5$ | $4.0\times10^5$ |

## 3. Discussion

3.1. It is clear that the $H_\beta$ measurements in Fig. 2 do not agree well with the result of numerical calculated by formula (1). The measurements are non-monotonic from center to limb (first decreasing and then increasing) while the line-of-sight integral of electron density monotonically increases from center to limb.

Chae et al. (1977) show that if there are a large number of tiny magnetic flux tubes, which are embedded vertically in the solar atmosphere with both upflows and downflows, the downflows should be brighter by one or two orders of magnitude because of the differences in the energy balance. These phenomena will also cause the solar lines' redshift, but the resultant redshift at the center will be larger than at the limb. Combining the effect proposed by Chae et al. (1977) with formula (1), the measurements can be fitted well. The result of fitting is shown in Fig. 3.

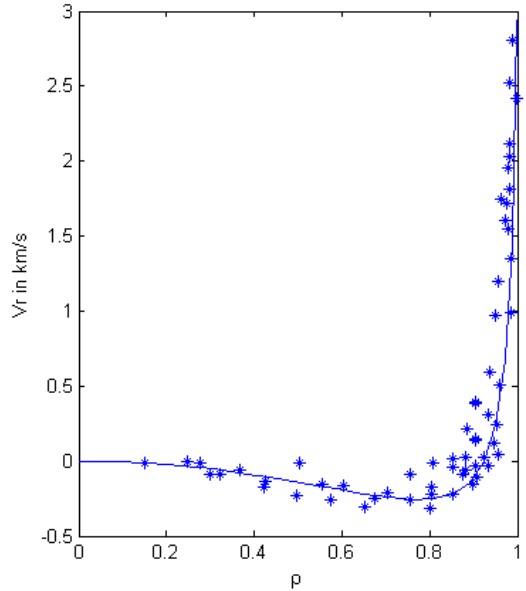

Fig3. The observed solar limb effect and the combined result of two effects

3.2. Formula (1) is derived under the assumption that the region is full of dilute plasma. In the solar corona this assumption is correct. In the solar chromosphere this assumption may have some problem. From Table 1 it can be seen that under h < 1900 km region the material is not fully ionized. Therefore, the applicability of formula (1) may be questionable. Numerical calculation shows that the factor $\frac{2e^2}{\pi.\eta}\frac{\ln \Lambda}{V_e}$ in the corona is almost a constant. Under h < 1900 km region the factor $\frac{2e^2}{\pi.\eta}\frac{\ln \Lambda}{V_e}$ increases rapidly as the height h decreases. Fig. 4 shows the variation of this factor.

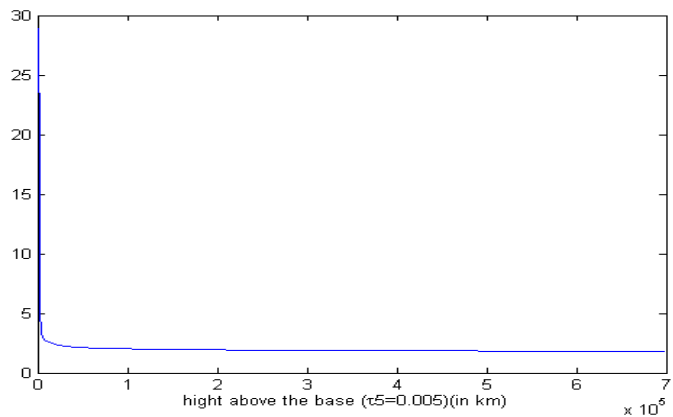

Fig4. The variation of the factor $\frac{2e^2}{\pi.\eta}\frac{\ln \Lambda}{V_e}$ in the solar atmosphere

3.3. The observed center-to-limb wavelength variations for different lines are varying differently (Kierin and Sharp 1968, Adam et al. 1976). This phenomenon is very easy to understand according to formula (1). The different lines are formed in different regions in the solar atmosphere. For lines which are formed at higher layers the variations of wavelength calculated by formula (1) should be smaller.

**Acknowledgement**

I would like to thank Dr. Nailong Wu for the correction and suggestions he made to greatly improve the English of the manuscript.